	\documentclass[a4paper,fleqn]{cas-dc}

	\usepackage[authoryear,longnamesfirst]{natbib}
	\usepackage[hyphenbreaks]{breakurl}

\begin{document}
	\let\WriteBookmarks\relax
	\def\floatpagepagefraction{1}
	\def\textpagefraction{.001}
	
	\shorttitle{Vector Centrality in Hypergraphs}
	
	\shortauthors{K Kovalenko et~al.}
	
	\title [mode = title]{Vector Centrality in Hypergraphs}                      

	\author[1]{K. Kovalenko}
	\fnmark[1]
	\author[2]{M. Romance}
	\fnmark[1]
	\author[1,3]{E. Vasilyeva}
	\fnmark[1]
	\author[2]{D. Aleja}
	\author[2]{R. Criado}
	\author[1,4,5]{D. Musatov}
	\author[1,4,6,7]{A.M. Raigorodskii}
	\author[2]{J. Flores}
	\author[8]{I. Samoylenko}
	\author[2]{K. Alfaro-Bittner}
	\cormark[1]
	\ead{karin.alfaro@urjc.es}
	\author[9,10,11,12]{M. Perc}
	\author[1,2,13]{S. Boccaletti}
	
	\address[1]{Moscow Institute of Physics and Technology, 9 Institutskiy per., Dolgoprudny, 141701, Russia}
	
	\address[2]{Universidad Rey Juan Carlos, Calle Tulip\'an s/n, 28933 M\'ostoles, Madrid, Spain}
	
	\address[3]{P.N. Lebedev Physical Institute of the Russian Academy of Sciences, 53 Leninsky prosp., 119991 Moscow, Russia}
	
	\address[4]{Caucasus Mathematical Center at Adyghe State University, ul. Pervomaiskaya, 208, Maykop, 385000, Russia}
	
	\address[5]{Russian Academy of National Economy and Public Administration, pr. Vernadskogo, 84, Moscow, 119606, Russia}
	
	\address[6]{Mechanics and Mathematics Faculty, Moscow State University, Leninskie Gory, 1, Moscow, 119991, Russia}
	
	\address[7]{Institute of Mathematics and Computer Science, Buryat State University, ul. Ranzhurova, 5, Ulan-Ude, 670000, Russia}
	
	\address[8]{National Research University Higher School of Economics, 6 Usacheva str., Moscow, 119048, Russia}
	
	\address[9]{Faculty of Natural Sciences and Mathematics, University of Maribor, Koro{\v s}ka cesta 160, 2000 Maribor, Slovenia}
	
	\address[10]{Department of Medical Research, China Medical University Hospital, China Medical University, Taichung 404332, Taiwan}
	
	\address[11]{Complexity Science Hub Vienna, Josefst{\"a}dterstra{\ss}e 39, 1080 Vienna, Austria}
	
	\address[12]{Alma Mater Europaea, Slovenska ulica 17, 2000 Maribor, Slovenia}
	
	\address[13]{CNR - Institute of Complex Systems, Via Madonna del Piano 10, I-50019 Sesto Fiorentino, Italy}

	\cortext[cor1]{Corresponding author}
	
	\fntext[fn1]{These three authors equally contributed to the Manuscript.}

	\begin{abstract}
	Identifying the most influential nodes in networked systems is of vital importance to optimize their function and control. Several scalar metrics have been proposed to that effect, but the recent shift in focus towards network structures which go beyond a simple collection of dyadic interactions has rendered them void of performance guarantees. We here introduce a new measure of node's centrality, which is no longer a scalar value, but a vector with dimension one lower than the highest order of interaction in a hypergraph. Such a vectorial measure is linked to the eigenvector centrality for networks containing only dyadic interactions, but it has a significant added value in all other situations where interactions occur at higher-orders. In particular, it is able to unveil different roles which may be played by the same node at different orders of interactions -- information that is otherwise impossible to retrieve by single scalar measures. We demonstrate the efficacy of our measure with applications to synthetic networks and to three real world hypergraphs, and compare our results with those obtained by applying other scalar measures of centrality proposed in the literature.
	\end{abstract}
	
	
	

\maketitle
	
\section{Introduction}
	
Ranking nodes in a graph is certainly the most fundamental task in modern network science~\cite{boccaletti_pr06, newmann10, jackson2010social, barabasi16}. Already in 1977, Linton C. Freeman gave the first definition of betweenness centrality, and used it to rank individual clout in social networks~\cite{freeman1977set, freeman1978centrality}. The earliest definition and use of eigenvector centrality can even be traced more than a century ago, in 1895, when Edmund Landau used it for scoring chess tournaments~\cite{landau1895relativen}. Nonetheless, it was not before the discovery of heterogeneity in the degree distributions of real world networks~\cite{barabasi_s99} that the full depth of implications of node centrality was realized. The `hub' became, and still is, a popular meme that stands for influence, importance, or virality  in social, biological and technological networks~\cite{albert_rmp02, newman_siamr03, boccaletti_pr06, barthelemy_pr11, holme_pr12}. The identification of the most central nodes in complex networks is crucial for error and attack tolerance~\cite{albert2000error, cohen2001breakdown}, viral marketing~\cite{richardson2002mining}, information spreading~\cite{kitsak2010identification, kleinberg1999authoritative, van2013network}, influence maximization~\cite{morone_n15, lu2016vital}, as well as plant genomic \cite{parenclitic1} and cancer research \cite{parenclitic2, parenclitic3}, just to name but a few examples.  Not to mention that companies like Google are actually building their entire business in providing efficient and customized rankings of webpages.
	
Although the relevance of quantifying node centrality is undisputed, the best measure for it very much depends on the particularities of the problem at hand. The various measures adopted so far to quantify node centrality, from the simplest node degree to the variations of betweenness and eigenvector centrality~\cite{brandes2001faster, newman2005measure, piraveenan2013percolation, alvarez2015eigencentrality}, do not optimize a global function of influence, and are thus inherently unable to guarantee optimal performance~\cite{morone_n15}. Therefore, the correct question one has to ask himself is not how central is a given node in a network, but rather how central is a given node in a network {\it with respect to a given process}.

\sloppy	
The issue is further exacerbated by the recent departures from traditional networks towards multilayer and higher-order networks as more apt representations of real world systems~\cite{boccaletti_pr14, kivela_jcn14, battiston2020networks}. Although a generalization of eigenvector centrality for multiplex networks has been proposed~\cite{sola2013eigenvector}, this does not account for the fact that in higher-order networks a link can connect more than two nodes. The potential of higher-order interactions has been recognized already in the early 70s by Ronald H. Atkin~\cite{atkin1972cohomology}, but the interest peaked only recently with the inability of classic graph representations to describe group interactions. This ineptitude comes to a head when studying peer pressure, public cooperation, complex contagion or opinion formation, to list just a few examples that clearly extend well beyond dyadic interactions in social science~\cite{burgio2020evolution, alvarez-rodriguez_nhb21}, or when considering three or more species that routinely compete for food and territory in a complex ecosystem~\cite{levine2017}, or when functional~\cite{lord2016} or structural~\cite{sizemore2018} brain networks or protein interaction networks~\cite{estrada2018} are studied.
Several different approaches to define higher-order interactions have been considered in the literature  (see the review works~\cite{battiston2020networks,ref31,ref32,ref33} for a comprehensive account of the different definitions adopted for higher-order interactions).
In specific instances, higher-order interactions have been modeled as (directed) sequences of nodes, sequences of set interactions, or motifs of dyadic edges, or combinations of two or several of those types~\cite{ref31,ref32,ref33}. In our work, we adopt what is possibly the most common notion of higher-order networks, i.e., is that of hypergraphs or simplicial complexes, where interactions between nodes are represented by a generalization of edges to hyperedges which capture (undirected and unweighted) group interactions. The generalization of our study to other settings will be considered in future works, in the line of what was recently suggested in Refs.~\cite{ref34,ref35,ref36}.
	
\begin{figure}
	\centering
	\includegraphics[width=8.0cm]{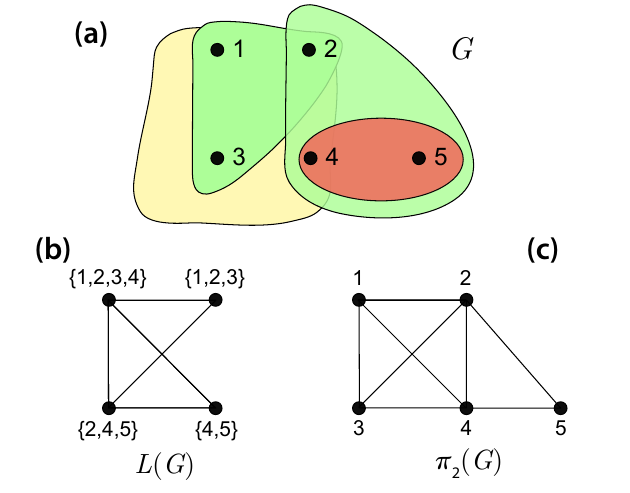}
	\caption{An illustrative example of a linegraph $L(G)$ [panel (b)] of a higher-order network $G=(V,\mathcal{E})$ with five nodes [panel (a)] and its projection network $\pi_2(G)$ [panel (c)]. See text for specifications.}
\label{fig:example}
\end{figure}
	
In view of these recent developments, it is therefore crucial to generalize centrality measures in a way that they can account for higher-order interactions. In fact, some measures have been introduced in the literature that extends the classic notion of centrality to hypernetworks~\cite{estrada2006,benson2019three,serrano2020centrality,tudisco2021node}, but they all compute a single (scalar) number per node.
In our study, we consider the most general case of an ensemble of $N$ nodes which interplay by means of interactions of any order $d \leq D$ (with $D$ indicating the maximum order of group interactions taking place in the ensemble), and introduce instead a  measure of centrality which is {\it a vector} assigned to each node, with dimension $D-1$. While our vector centrality is related to the classical eigenvector centrality for networks containing only dyadic interactions,  we will demonstrate that our  measure has instead a significant added value (if compared with scalar measures) in all situations where interactions occur at higher-orders. We will show with practical applications that our measure is, indeed, able to distinguish different roles which may be played by a same node at different orders of interactions, a feature which is evidently impossible to be revealed by any single scalar measure.

{\bf The vector centrality measure. }Let us start with considering $N$ nodes which are interplaying by means of $l_2$ links (dyadic interactions), $l_3$ hyperlinks of order 3 (triadic interactions), $l_4$ hyperlinks of order 4 (quadratic interactions), and in general by $l_d$ hyperlinks of order $d$ (with $d=2,3,...,D$). We here concentrate on the case where all such hyperlinks are undirected. Mathematically this defines an undirected higher-order network (or hypergraph)  $G=(V, \mathcal{E})$, i.e., a finite set $V$ containing $N$ nodes, and a family $\mathcal{E}$ of $\mathcal{\ell}=\sum_{i=2}^{D} l_i$ non-empty and non-singleton subsets of nodes of $G$, each subset defining a hyperlink.
	
Our idea is to associate to $G$ its linegraph $L(G)$, as introduced by Hassler Whitney for graphs in 1932 \cite{Whitney1932} and extended for higher-order networks by Jean-Claude Bermond et al. in 1977~\cite{Bermond1977, Heydemann1978}. In particular, $L(G)$ is a graph of $\mathcal{\ell}$ nodes (each of which mapping one of the hyperedges of $G$). The links of $L(G)$ stand for adjacency between hyperedges in $G$: if $h_1 \in \mathcal{E}$ and $h_2 \in \mathcal{E}$ are two hyperlinks, then there is an undirected link in $L(G)$ between the nodes $h_1$ and $h_2$ if and only if $h_1\cap h_2\ne \emptyset$.
	
Figure~\ref{fig:example} depicts an illustrative example, where a hypergraph $G=(V, \mathcal{E})$ is defined by $V=\{1,2,3,4,5\}$ and $\mathcal{E}=\{\{1,2,3\},\{1,2,3,4\},\{2,4,5\},\{4,5\}\}$. The figure shows also the associated linegraph $L(G)$, and the projection network $\pi_2(G)$ of $G$. Notice that the projection of an hypergraph into a graph can be constructed in different ways. For instance, it can be defined as an unweighted network (see Ref.~\cite{battiston2020networks}) given by the number of hyperedges between two nodes \cite{Carletti2020}, or as a weighted network~\cite{Banerjee2017}, among others formalisms. In our work, the projection network $\pi_2(G)$ of the hypergraph $G=(V, \mathcal{E})$  is defined as the classic undirected and unweighted network formed by the same set of nodes as in $G$ and whose links represent the dyadic interactions resulting from the projection of the hyperlinks of $G$.

Now, it is straightforward to demonstrate that if $G$ is undirected and connected, then also $L(G)$ is undirected and connected. Indeed, for any pair of hyperedges $h_i$ and $h_j$ in $L(G)$ a path can be constructed by choosing a node $v$ from $h_i$ and a node $w$ from $h_j$, and by using the same sequence of hyperedges as in path from $v$ to $w$ in $G$.
Then, the classic Perron-Frobenius theorem \cite{Perron_1907theorie,frobenius1912matrizen} guarantees the existence and uniqueness of the eigenvector centrality of $L(G)$.
In other words, one can compute with standard methods the classical eigenvector centrality of each node in $L(G)$, and one obtains a value $c(h)\in[0,1]$ for all  hyperlinks $h\in\mathcal{E}$ in $G$, such that $\sum_{h\in\mathcal{E}}c(h)=1$.
	
With the $\mathcal{\ell}$ values of $c(h)$ at hand, we can now define the {\it vector centrality} of each node $i\in V$, a non-negative vector $\vec{c_i}=(c_{i2},\cdots,c_{iD})\in \mathbb{R}^{D-1}$ such that, for every $2\le k\le D$ one has
\begin{equation}\label{def:vectorial}
	c_{ik}=\frac 1k\sum_{\substack{i\in h\in \mathcal{E} \\ |h|=k}}c(h),
\end{equation}
where $|h|$ indicates the order (or size) of the hyperedge $h$, and $D=\max\{|h|;\enspace h\in\mathcal{E}\}$ is the maximal size of hyperedges in $G$ (the maximal order of the group interactions affecting the $N$ nodes in the ensemble).

In other words, the $k^{th}$ component $c_{ik}$ of the vector centrality of node $i$ is the sum of the centralities of all hyperlinks of size $k$ that contain $i$ as one of the incident nodes, and the weight value $\frac 1k$ makes that
	
\begin{equation}\label{eq:normalization}
	\sum_{i\in V}\|\vec{c_i}\|_1=\sum_{i\in V}\sum_{k=2}^Dc_{ik}=\sum_{i\in V}\sum_{k=2}^D\sum_{\substack{i\in h\in \mathcal{E} \\ |h|=k}}\frac {c(h)}k .
\end{equation}
Now, for each $i\in V$, one has that
\[
\sum_{k=2}^D\sum_{\substack{i\in h\in \mathcal{E} \\ |h|=k}}\frac {c(h)}k=\sum_{i\in h\in\mathcal{E}}\frac{c(h)}{|h|}
\]
because in the last double summation each hyperlink $h\in \mathcal{E}$ such that $i\in h$ appears exactly once.
Therefore, by using this last expression in Eq.~(\ref{eq:normalization}), and by summing over all nodes $i\in V$,  one gets that
\[
\begin{split}
	\sum_{i\in V}\|\vec{c_i}\|_1&=\sum_{i\in V}\sum_{i\in h\in\mathcal{E}}\frac{c(h)}{|h|}\\
	&=\sum_{h\in\mathcal{E}}\sum_{i\in h}\frac{c(h)}{|h|}=\sum_{h\in\mathcal{E}}c(h)=1.
	\end{split}
\]
This latter expression has been obtained by simply changing the summation order, and taking into account that every summand $\frac{c(h)}{|h|}$ appears exactly $|h|$ times. The final result is, therefore, that $\sum_{i\in V}\|\vec{c_i}\|_1=1$, which implies that our measure is	properly normalized.
	
Notice that if $D=2$, i.e., only dyadic interactions exist in $G$, then for each node $i$ one has $\vec{c_i}=(c_{i2})\in\mathbb{R}$, where the scalar value $c_{i2}$ is related with the $i^{th}$ component ($c_i'$) of the classic eigenvector centrality of $G$, as it was proved in~\cite{CFGR2011}.  Precisely, for $D=2$, calling $\lambda_1$ and $\lambda_2$ the greatest and second greatest eigenvalue of the adjacency matrix of $G$, and denoting by $\Delta$ the norm of the difference between our measure and the eigenvector centrality, Ref.~\cite{CFGR2014} gave the following bounding relationship:
\begin{equation*}
	\Delta \equiv \sqrt{\sum_{i=1}^{N}(c_{i2}-c_i')^2}\leq\frac{(4-\sqrt{2})\sqrt[4]{2}\sqrt{N}\sqrt[8]{2 l_2}}{\lambda_1-\lambda_2}\sqrt[4]{\lambda_1-\frac{2 l_2}{N}},
\end{equation*}
which holds as far as the so called graph irregularity $I(G)=\lambda_1-\frac{2 l_2}{N}$ is smaller than $((\sqrt{2}-1)^2/(4\sqrt{2}-2)^4N^2\sqrt{2 l_2})(\lambda_1-\lambda_2)^4$ (see Ref.~\cite{CFGR2014} for details). We remark that this expression shows analytically that, for $D=2$, i.e., when only dyadic interactions exist in $G$,  $c_{i2}$ is very close to the $i^{th}$ component ($c_i'$) of the classic eigenvector centrality of $G$, and actually the more regular the network is the closer $c_{i2}$ is to $c_i'$. In fact, it can actually be demonstrated that a more complicated construction of $L(G)$ (where the setting of both nodes and links would account also for all possible permutations in the order of the nodes forming hyperlinks in $G$), would lead to recover exactly (at $D=2$) the classical eigenvector centrality. For all practical purposes of this article, however, such a ``directed" construction leads to the same qualitative results and quantitative rankings, and therefore we decided to report it elsewhere as a mathematical extension of our measure.
	
Moreover, we would like to remark here that the same entire procedure can be actually used for extensions of other structural measure (like, for instance, node betweenness) to higher-order networks: given a hypergraph $G$, one can always construct the associated linegraph, calculate the measure values for all hyperedges, and then use expressions similar to~(\ref{def:vectorial}) to define vectorial quantities associated to the nodes in $G$.
	
\begin{figure*}
	\centering
	\includegraphics[scale=0.90]{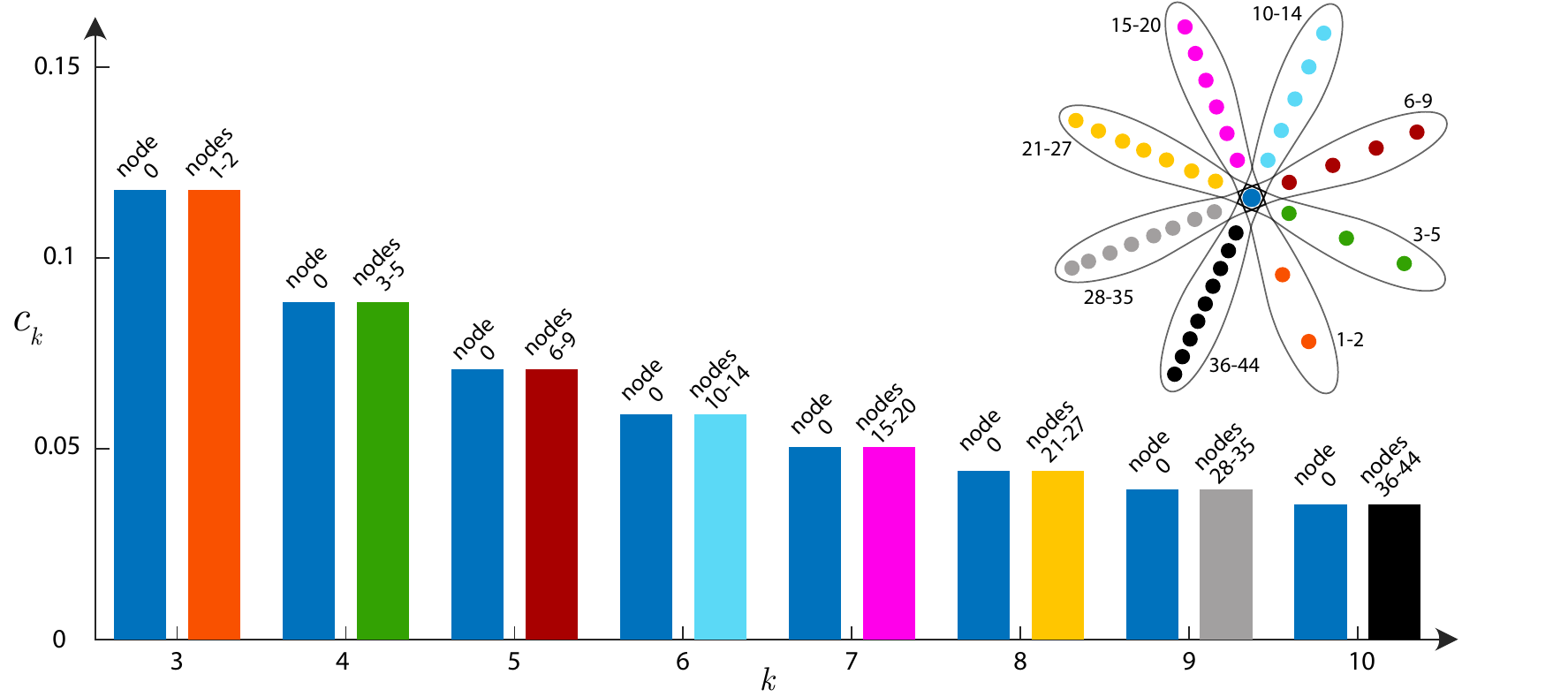}
	\caption{The values of all components of the vector centrality for the nodes of a sunflower hypergraph with eight petals, each one corresponding to a hyperedge of different size (from 3 to 10, as it is seen in the pictorial sketch at the right of the figure, which has to be regarded also for the color code of the different bars appearing in the main plot). 	Our measure allows to clearly distinguish the properties of the central node's  from those of all other nodes, as node 0 it is the only one having non-zero values in all its components, whereas the centrality of all other nodes is {\it localized} only in the component corresponding to the order of the hyperedge to which they are belonging.}
\label{sunflower}
\end{figure*}

\section{Results}
	
We here consider several practical examples to illustrate the added value of our vectorial centrality in distinguishing different roles a given node may have with respect to processes which may occur on top of interactions of different orders, a capacity which is instead greatly dwindled, if not prevented at all, using classical measures on the (weighted or unweighted) projections of the hypergraph.

	
{\bf Synthetic networks. }In order to provide a first comparison between our vectorial measure and the scalar centralities proposed so far, 	let us refer to the measures introduced in Refs.~\cite{estrada2006,benson2019three,serrano2020centrality,tudisco2021node}.
	
Reference~\cite{serrano2020centrality} extends the methodology of  eigenvector centrality for the case of simplicial complexes. A direct comparison with our measure is therefore not possible, as in simplicial complex one has to assume that the existence of a $d$-simplex (a simplex of order $d$) automatically implies the existence of all possible interaction orders from 2 to $d-1$, which is not the case for many real world higher-order networks, and which makes that framework totally different from the more general case of hyper-networks considered here.
	
In Ref.~\cite{benson2019three} three possible generalizations of eigenvector centrality for regular hypergraph are presented. The evident difference between our measure and the ones proposed in~\cite{benson2019three} is therefore the fact that we do not restrict hypergraph to be regular. However, even if we limit ourselves to the case of regular hypergraph, our results differ from the ones presented in~\cite{benson2019three}. In particular, let us refer to the same example that was made in  Ref.~\cite{benson2019three}, the so called sunflower hypergraph, a star-like hypergraph having one central node and $r$ edges (sunflower's petals), each one of order $d$. In this example, our measure directly provides the ratio between the value of centrality of the central node, $c_{0d}$,  and that of every other node $i$, $c_{id}$. One indeed has that $c_{0d} = \frac{1}{d}\sum_{0\in h\in \mathcal{E},|h|=d}c(h) = \frac{rc}{d}$ and $c_{id} = \frac{1}{d}\sum_{i\in h\in \mathcal{E},|h|=d}c(h) = \frac{c}{d}, \ i\ne 0$, with $c$ being the line graph nodes' centrality (which, in this case, is equal for each node in $L(G)$, as the line graph is a clique of $r$ nodes). It immediately follows that $c_{0d}/c_{id}=r$, which makes a strong difference with respect to what reported in Fig.~1 of Ref.~\cite{benson2019three}.
	
Finally, Ref.~\cite{tudisco2021node} suggests a generalization of the famous HITS algorithm to hypergraphs, in which the nodes are more central if they are connected with more central hyperedges, and vice versa. This approach shares, indeed, similarities with the idea proposed in our study, as the components of our vector centrality are calculated from the values of the hyperedges' centralities in the line graph.
However, the information that one can extract from the two measurements is completely different. For instance, let us analyze the same example presented in Ref.~\cite{tudisco2021node}, i.e., a sunflower hypergraph with eight petals, which however correspond now to hyperedges of different sizes (from 3 to 10, see the pictorial sketch at the left of Fig.~\ref{sunflower}). In Ref.~\cite{tudisco2021node} node 0 has the highest centrality value, no matter which function (linear, max, log-exp) is used in the process of centrality calculation. For all other nodes, in the linear case (the log-exp case) the higher is the order of the hyperedge to which they belong the higher (the lower) is the value of the centrality, while in the max case the centrality values are all equal.
In comparison, our vectorial measure provides a much richer information, as one can immediately see from Fig.~\ref{sunflower}.
The central node's properties are now clearly distinguishable from those of all other nodes, primarily because it is the only one having non-zero values in all its centrality's components, whereas all other nodes feature a {\it localized} centrality value in the component corresponding to the order of the hyperedge to which they are belonging. Furthermore, at each hypergraph order, all the nodes in the corresponding petal have the same centrality component, once again differentiating our results from those of Ref.~\cite{tudisco2021node}.
	
In order to better illustrate the qualities of our vectorial measure, let us now move to a more complicated synthetic hypergraph consisting of $N=100$ nodes, 400 hyperlinks of order 2, 400 hyperlinks of order 3 and 400 hyperlinks of order 4 (i.e., $l_2=l_3=l_4=400$), mapping therefore an ensemble of units interplaying by means of dyadic, triadic, and quadratic interactions. Here, we want to highlight how our vectorial centrality outperforms classical measures in tracking the importance of nodes when changes occur in the network structure. To this purpose, we initially prepare a graph with all $l_3$ hyperlinks of order 3 which are randomly distributed.  As instead for the $l_2$ links of order 2 (the $l_4$ hyperlinks of order 4), $350$ of them are distributed randomly, whereas $50$ of them are placed so as to make vertex 1 (vertex 100) a hub  for dyadic (quadratic) interactions, i.e., they are constructed so as to include vertex 1 (vertex 100) as one of the incident nodes. Then, we simulate limitations processes in group interactions by removing at random a fraction $p$ of quadratic interactions, and we keep track on how the different centrality measures are efficient in monitoring the change of relevance of each node following the changes in the network structure. Precisely, one surveils the behavior of $c_{1,2}$ (the first component of the vector centrality of node 1), of $c_{N,4}$ (the last component of the vector centrality of node 100), $c_{1,\pi}$ and $c_{N,\pi}$ (the first and last components of the classical eigenvector centrality for the projected graph, $\vec{c}_\pi$), and $c_{1,\pi_w}$ and $c_{N,\pi_w}$ (the first and last components of the classical eigenvector centrality for the projected weighted graph, $\vec{c}_{\pi_w}$). In $\pi$ two nodes are connected if there exists a hyper-link to which they both belong to; in ${\pi_w}$ the weight of each link is the number of hyperlinks to which the two end nodes are belonging to.
	
	\begin{figure}
		\centering
		\includegraphics[scale=1.2]{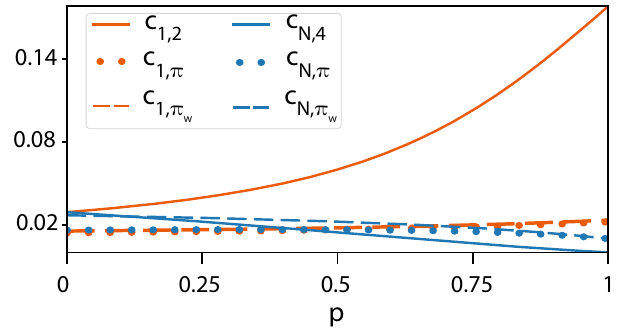}
		\caption{$c_{1,2}$, $c_{N,4}$, $c_{1,\pi}$, $c_{N,\pi}$, $c_{1,\pi_w}$ and $c_{N,\pi_w}$ (see text for definition) vs. the fraction $p$ of removed quadratic interactions, for the first synthetic network described in the text.
			The color code of the different curves is reported in the legend. Each point corresponds to an ensemble average over 5,000 simulations: 100 different network realizations and for each one of them 50 different realizations of random removal of the $l_4$ hyperlinks.}
		\label{fig:synthetic}
	\end{figure}

The results are reported in Fig.~\ref{fig:synthetic}, and show clearly that only our vectorial measure (by comparison of $c_{1,2}$ and $c_{N,4}$) is able to reveal a substantial loss of centrality of node 100 as the number of quadratic interactions is progressively reduced, and a corresponding gain in centrality of node 2 which eventually remains the only hub in the system.
	
In a second example of a synthetic network, we probe the capability of our measure to reveal different scaling properties which may affect different orders of interactions in the graph, even in the case in which, at variance with the previous case, such orders do not correspond to the same number of hyperlinks. To that purpose, we construct another synthetic graph with dyadic, triadic, and quadratic interactions, this time with $N=1,000$, $l_2= 4,000$ hyperlinks of order 2, $l_3=1,000$ hyperlinks of order 3, and $l_4=2,000$ hyperlinks of order 4.
All $l_2$ and $l_4$ hyperlinks are chosen randomly, this way determining a strongly homogeneous distribution for dyadic and quadratic interactions.
Instead, the $l_3$ hyperlinks (which are in the minimum number with respect to all other hyperlinks) are chosen so as to determine a strongly heterogeneous distribution: at each time those hyperlinks are constructed with a probability which explicitly depends on the actual node degree.
	
It has to be remarked that any projection (weighted or unweighted) of such synthetic higher-order network would result in a heterogeneous degree distribution, with the consequence that any centrality measure applied to such projected graph would reveal a strong heterogeneity. The results of applying our measure are, instead, shown in Fig.~\ref{fig:synthetic1}, where we report the histograms of the first ($c_2$, panel a), the second ($c_3$, panel b) and the third ($c_4$, panel c) component of our vectorial centrality. It is seen that while the histograms reveal strong homogeneity at the level of dyadic and quadratic interactions (panels a and c), they clearly show heterogeneity traits at the level of triadic interactions (panel b), this way accounting exhaustively for the overall structural properties which have been engineered in the hypergraph.


{\bf Real-world hypergraphs. }Finally, we calculate our measure on several real-world hypergraphs and discuss the added value of our measure in revealing important information on the structure of the considered hypernetworks.
	
The first considered hypergraph is that mapping the information publicly available in  the arXiv \sloppy ({\url{https://arxiv.org/}, \url{https://github.com/mattbierbaum/arxiv-public-datasets/}) database, with the data parsing made by Ref.~\cite{clement2019use}. In particular, we focus on the data of preprints published in mathematics, and extract those papers which were written in collaboration, i.e., those having at least two co-authors. The extracted dataset consists of a total of 498,071 papers 	co-written by 230,605 authors.
		
\begin{figure}
	\centering
	\includegraphics[scale=1.2]{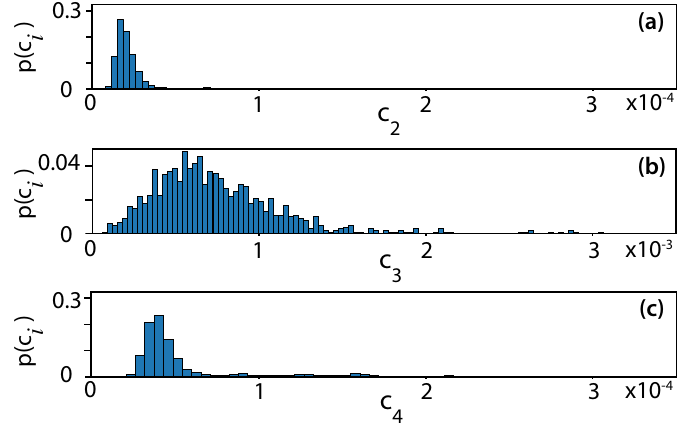}
	\caption{(a-c) Histograms (sampled with 100 bins) of the first ($c_2$, panel a), the second ($c_3$, panel b) and the third ($c_4$, panel c) component of the vectorial centrality, calculated over the second synthetic network described in the main text. Each histogram refers to values which are ensemble averaged over 10 different graph realizations.}
\label{fig:synthetic1}
\end{figure}
		
The data were mapped into a hypergraph $G_{math}$, where nodes were scientists, and each paper formed a hyperlink (a group interaction) of length equal to the number of co-authors. The maximal number of co-authors of a single papers (i.e., the maximal length $D$ of hyperlinks in $G_{math}$) is 67, which implies that the vectorial centrality of each scientist will have 66 components.
The associated linegraph $L(G_{math})$ is rather large in size: it is obviously formed by 498,071 nodes, and it has 9,808,188 links.
The eigenvector centrality of $L(G_{math})$ is then calculated, and the vector centrality of each scientist in $G_{math}$ is evaluated.
		
Various rankings of scientists may be extracted according to the different components of the vector centrality, i.e., scientists may have different	role and importance with respect to different hyperedges' sizes. In particular, we here analyze how many of the members of the top $x$ authors's list in the ranking with respect to a given component of the vector centrality is also belonging to the top $x$ authors's list in the ranking made with respect to another component. To do so, we introduce the fraction $\mu_x$ as follows:
		
\begin{equation}
	\label{mu_2}
	\mu_x(c^{i}, c^{j}) = \frac{|{\rm top}_x(c^i)\cap{\rm top}_x(c^j)|}{x},
\end{equation} 	
		
\noindent where $c^{i}$ and $c^{j}$ are, respectively, the $i^{th}$ and $j^{th}$ components of the vector centrality of the nodes, ${\rm top}_x(c^i)$ (${\rm top}_x(c^j)$) is the set of the nodes which are occupying the top $x$ positions in the ranking made by comparing the $i^{th}$ (the $j^{th}$) component of their vector centralities, and $|\cdot|$ stands here for the cardinality of the set. $\mu_x(c^{i}, c^{j})$  measures therefore how large is the overlap between the two sets, and its values $\mu_x(c^{i}, c^{j})$ form a square matrix of $66 \times 66$ elements, which actually describes how correlated are the positions scientists are holding in the ranking calculated with respect to a given component of the vector centrality with those held by the same scientists in the ranking calculated with respect to another component.
		
\begin{figure}
	\centering
	\includegraphics[scale=1.02]{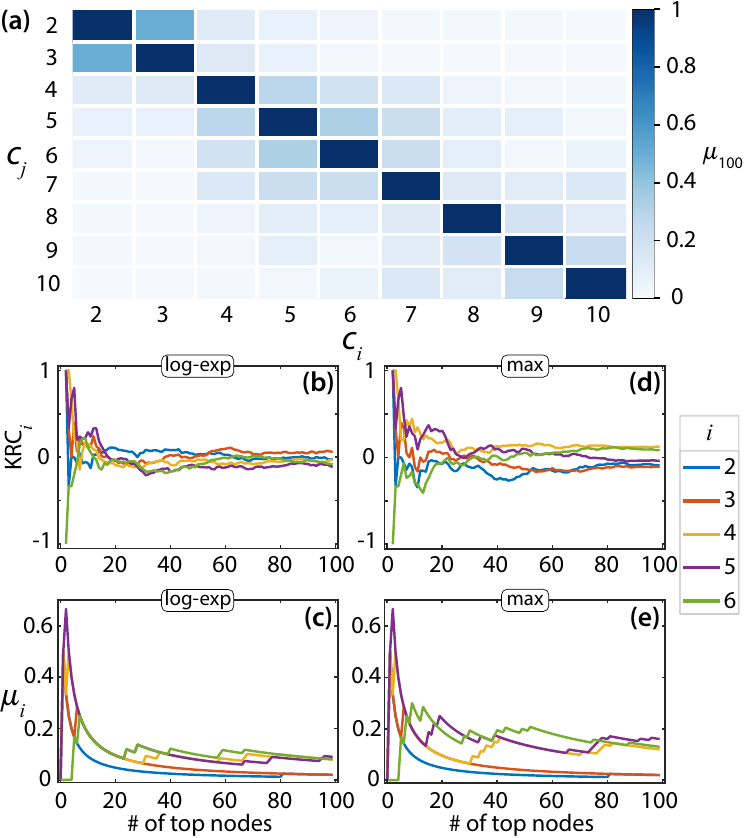}
	\caption{ (a) $\mu_{100}(c^{i}, c^{j})$ (see Eq.~(\ref{mu_2}) of the text for definition) for the hypergraph representing scientific co-authorship in  mathematics. Reported values are limited to the first ten components, out of the 66, of the vector centrality.  It is clearly seen that, in general, the values of $\mu_{100}$ are rather small for $i \neq j$. (b-e) Correlation between the rankings provided by the different components (from the second to the sixth, see color code in the legend at the right of the figure) of our vector centrality and the unique ranking obtained by adopting the algorithm of Ref.~\cite{tudisco2021node} with a log-exp (b,d) and a max (c,e) function. Panels b and c report the Kendall rank correlation (KRC) coefficients, while panels d and e report the values of the function $\mu$ (from the same Eq.~(\ref{mu_2}), in which $c^j$ are substituted with the values of centralities extracted with the algorithm of Ref.~\cite{tudisco2021node}). }
\label{corr}
\end{figure}
		
The values of $\mu_{100}(c^{i}, c^{j})$ (limited to the first ten components, out of the 66, of the vector centrality) are reported in panel a) of Fig.~\ref{corr}. It is seen that, except for the few values close to $i=j$, the fractions  $\mu_{100}(c^{i}, c^{j})$  are relatively small for $i \neq j$ and, therefore, the lists of the 100 top leaders in the rankings made with respect to different hyperedges sizes are significantly different. This confirms that the use of our measure is essential for extracting information on such differences, which would be instead unaccessible by any other scalar measure of centrality.
In panels (b-e) of Fig.~\ref{corr} a comparison is made with the unique ranking obtained by the use of the scalar measure of centrality proposed in Ref.~\cite{tudisco2021node} with a log-exp (panels b and d) and a max (panels c and e) function. This is done by reporting two different correlation measures: the Kendall rank correlation (KRC) coefficients (panels b and c) and the values of the function $\mu$ as calculated by  Eq.~(\ref{mu_2}) when the values $c^j$ are substituted with those extracted with the algorithm of Ref.~\cite{tudisco2021node} (panels d and e). Both correlation measures are reported as functions of the length of the ranking list. Lines of different colors correspond to different hyper-edge orders in our vector centrality measure. It is clearly seen that, when the ranking list is small in size, the intersections of the various sets of top ranked nodes is significant, implying that our measure individuates the same fundamental actor of the game. However, as the size of the ranking list increases, the corresponding Kendall rank correlation coefficients shrink, up to getting close to zero for every hyper-edge order. This implies that the obtained rankings do not differ substantially in individuating the really top nodes in the hypergraph, while they are fundamentally different as far as nodes of medium importance are considered.
\begin{figure}
	\centering
	\includegraphics[scale=1.02]{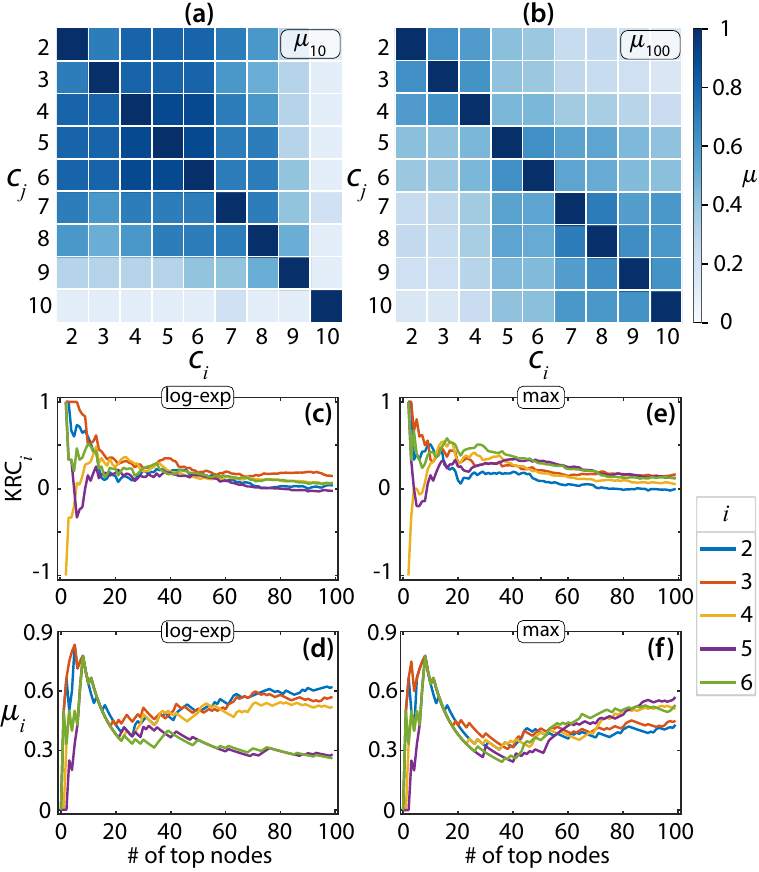}
	\caption{(a) $\mu_{10}(c^{i}, c^{j})$ and (b) $\mu_{100}(c^{i}, c^{j})$ (see Eq.~(\ref{mu_2}) for definition) for the commercial drug higher-order network (all specifications of the hypergraph are given in the text). (c-f) Correlation between the rankings provided by the different components (from the second to the sixth, see color code in the legend at the right of the figure) of our vector centrality and the unique ranking obtained by adopting the algorithm of Ref.~\cite{tudisco2021node} with a log-exp (c,e) and a max (d,f) function. Panels c and d report the KRC coefficients, while panels e and f report the values of the function $\mu$ (from the same Eq.~(\ref{mu_2}), in which $c^j$ are substituted with the values of centralities extracted with the algorithm of Ref.~\cite{tudisco2021node}).}
\label{drugandcontact}
\end{figure}

A second real-world hypergraph is constructed from the data available at \sloppy \url{https://www.cs.cornell.edu/~arb/data/NDC-substances/}~\cite{benson2018simplicial}. The dataset contains information on the composition of commercial drugs, posted by the U.S. Food and Drug Administration. In the dataset, each node represents a substance (or active principle, e.g., octinoxate, titanium dioxide, etc.) and each hyperlink stands for a commercial drug  made of a given composition of such active principles. For the purposes of our application, drugs composed by no more than 25 substances were taken (the same as in Ref.~\cite{benson2018simplicial}). Moreover, drugs consisting of only one active principle were excluded from the analysis. The result is an hypergraph consisting of 3,438 nodes and 29,296 hyperlinks.

Panels a) and b) of Fig.~\ref{drugandcontact} reports the results for $\mu_{10}$ and $\mu_{100}$, as defined by Eq.~\eqref{mu_2}.
The information that our measure provides allows to infer that there are principles which are important for both the drugs with small number of ingredients and the ones with complex composition [yielding a non negligible overlap between the top 10 ranked components for all hyperlink sizes in between 2 and 8, as can be seen in panel a) of  Fig.~\ref{drugandcontact}].  However, in general [see panel b) of Fig.~\ref{drugandcontact})] the sets of key ingredients of the drugs with simple and complex compound are significantly different. Once again, we remark that such information would have not been extracted from scalar centrality measures.
In panels (c-f) of Fig.~\ref{drugandcontact} we again compare the ranking obtained by the different components of our vector centrality measure with the unique ranking obtained by the use of the scalar measure of centrality proposed in Ref.~\cite{tudisco2021node} with a log-exp (panels c and e) and a max (panels d and f) function. One can actually draw the same conclusions as above: when the list is small in size, the top ranked nodes are common to both rankings, but when the size of the ranking list increases the KRC coefficients get close to zero for every hyper-edge order.

\begin{figure}
	\centering
	\includegraphics[scale=1.1]{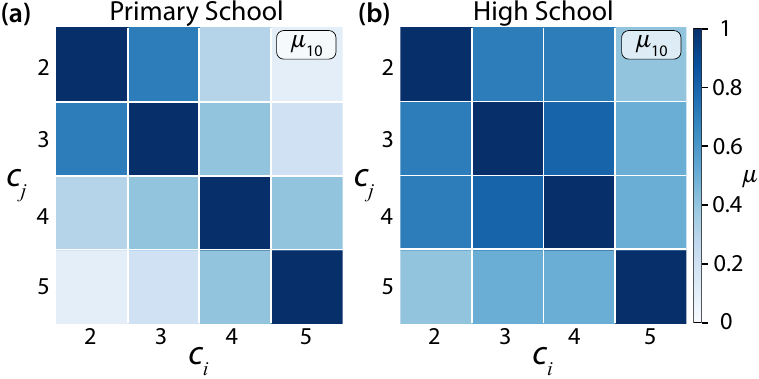}
	\caption{$\mu_{10}(c^{i}, c^{j})$ (see Eq.~(\ref{mu_2}) for definition) for the primary [panel (a)] and high [panel (b)] school contacts hypernetwork (all specifications of the hypergraph are described in the text). It is seen (with particular evidence in the case of high school contacts) that social interactions pilot the emergence of {\it leaderships} of students which tend to be central independently on the group size.}
\label{contact}
\end{figure}

As a third application, we consider the hypergraphs that can be constructed from the data on primary school contacts [taken from \sloppy \url{https://www.cs.cornell.edu/~arb/data/contact  -primary-school/} \cite{benson2018simplicial, Stehl2011contact}] and those on high school contacts [taken from \url{https://www.cs.cornell.edu/~arb/data/contact-high-school/}~\cite{benson2018simplicial, Mastrandrea2015contact}].
In both cases, data refer to experiments where wearable sensors, registering social interactions by proximity at a resolution of 20 seconds, are beard by students (242 kids in the case of the primary school, and 327 adolescents in the case of high school).  As the dataset contains a lot of repetitions of the same group of people (the duration of the interactions are in general far larger than the 20 second resolution time), only unique groups were analysed. Furthermore, only edges with size no less than 2 were considered.
The resulting hypergraph for primary school (high school) contacts consists therefore of 242 (327) nodes and 12,704 (7,818) hyperlinks corresponding to the total number of unique groups, i.e., the total number of nodes forming the line graph, which in its turn consists of a total number of edges of 2,238,167 (593,188).

For both hypergraphs, the largest group size (the maximal order of interaction) is 5. Once again, to compare the ranking of students related to distinct edge sizes, we use the same measure $\mu_x$, defined by Eq.~\eqref{mu_2}.
The results are shown in figure~\ref{contact}).
It is seen that {\it central} students in groups with size 2-4 are mainly not present in the top list of the groups of 5 people. However, it is seen (with particular evidence in the case of high school contacts) that such social interactions pilot the emergence of a {\it leadership} of students which tends to be central independently on the group size.
		
As the fourth application, we consider the hypergraph constructed from the data taken from
\url{https://www.cs.cornell.edu/~arb/data/senate-bills/} \cite{chodrow2021hypergraph, Fowler-2006-connecting, Fowler-2006-cosponsorship}.
There, nodes are US Congress persons and hyper-links are co-sponsorships of bills which were put forth in the Senate.
The dataset can be mapped into a hypergraph made of 294 nodes and 29,157 hyperedges, and the corresponding line graph has 29,157 nodes and 82,211,358 edges.
After application of our method, figure~\ref{wewillsee}	 reports the KRC coefficients (panel a, calculated now for each pair of ranking lists, as obtained with the $i^{th}$ and $j^{th}$ components of the vector centrality) and $\mu_{30}(c^{i}, c^{j})$  (panel b, limited to the first ten components of the centrality vectors).
From the plot $\mu_{30}(c^{i}, c^{j})$ it is rather evident that the leading roles are played always by the same actors (the parties' leaders), independently on the number of persons co-sponsoring the bill. Moreover, the orderings inside the leading groups are rather close to each other, as non-trivial KRC coefficients are obtained. Only the ranking with respect to hyperedges having size 2 seems to be weakly correlated with the others.

From the one hand this case is therefore rather different from all the previous ones, in that all components of the centrality vector are correlated rather strongly, and one could be tempted to say that there is no need here for the use of a vectorial measure. However, from the other hand it is only using our vector centrality that one can reveal that, in order to protect their leadership, central persons in political parties try to play key roles in groups of different sizes. In other words, a high correlation of the rankings related to different hyper-links orders gives also meaningful information.

Our final application is the hypergraph constructed with the data taken from \url{https://www.cs.cornell.edu/~arb/data/walmart-trips/} \cite{Amburg-2020-categorical}.
In this hypergraph, nodes are products at Walmart and hyperlinks are sets of co-purchased products. It is composed by 88,860 nodes and 69,906 hyperedges, with a corresponding line graph made of 69,906 nodes and 33,046,972 edges.
	
\begin{figure}
	\centering
	\includegraphics[scale=1.1]{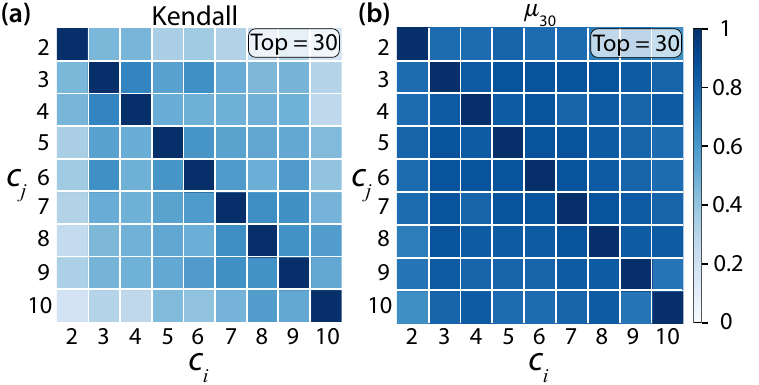}
	\caption{KRC coefficients (panel (a), calculated for each pair of ranking lists obtained with 				the $i^{th}$ and $j^{th}$ components of the vector centrality) and $\mu_{30}(c^{i}, c^{j})$  (panel (b), limited to the first ten components of the centrality vectors) for the hypergraph reflecting bills' co-sponsorships in the US Senate (all specifications of the hypergraph are described in the text).}
\label{wewillsee}
\end{figure}
	
The resulting KRC coefficients (panel a) and the values of $\mu_{10}(c^{i}, c^{j})$ (panel b) are presented in figure~\ref{wewillsee2}.
It is seen that the intersections of the sets of top 10 products with respect to different orders are very high. This means that there are some essential products which appear in each bill, no matter its sizes. However, when we analyze the set of top 100 products the sizes of these intersections progressively decrease, and the values of the KRC coefficients for vectors relating to the top 100 rankings are negligible (not shown in the figure). One can conclude that there exists a set of essential products which are bought frequently no matter which size the bill has. Other products are bought not to so frequently, and their appearance in the bill is not highly determined by the bill size. Once again, we highlight that such kind of conclusions can be drawn only when the centrality measure has a vectorial character.

\begin{figure}
	\centering
	\includegraphics[scale=1.1]{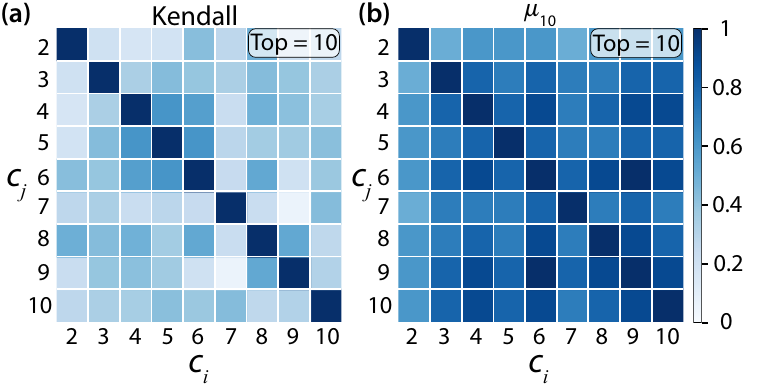}
	\caption{KRC coefficients (panel (a), calculated for each pair of ranking lists obtained with the $i^{th}$ and $j^{th}$ components of the vector centrality) and $\mu_{10}(c^{i}, c^{j})$  (panel (b), limited to the first ten components of the centrality vectors) for the hypergraph constructed from the Walmart-tips dataset (all specifications of the hypergraph are described in the text).}
	\label{wewillsee2}
\end{figure}
	
\section{Discussion}
	
Taken together, we have introduced a centrality measure able to overcome the inherent limitations of scalar centralities in higher-order networks. Our measure assigns a vector to each node, with dimension one lower than the dimension of the longest hyperlink in the network, and with every component thus determining the centrality of that node for a link with a particular length.
	
Our vector centrality is related to the classical eigenvector centrality for networks containing only dyadic interactions.
Furthermore, by using artificially generated higher-order networks as well as data from real-world higher-order networks, we have demonstrated that our  measure has instead a significant added value in all situations where interactions occur at higher-orders, in that it unveils different roles which may be played by a same node at different orders of interactions and therefore is the only one which accounts exhaustively for the properties of the overall interactive structure of the hypergraph.
In particular, our measure gives a much richer information about centrality relationships than that extracted from other scalar measures recently introduced for hypergraphs.
	
As noted already when introducing our vector centrality, the same approach can be readily applied to other structural measures, which thus opens the path towards a wider applicability of our approach.
	
We expect our  measure to become widely used with further progress in network science and related research fields.
	
\section{Acknowledgments} 
	
This work was supported by the Russian Federation Government (project "Post-crisis world order: challenges and technologies, competition and cooperation" funded by the  Ministry of Science and Higher Education, agreement number 075-15-2020-783), by the program ”Leading Scientific Schools ” (Grant No. NSh-775.2022.1.1), by the Spanish Government (Project PGC2018-101625-B-I00 (AEI/FEDER, UE)), by the URJC Grant No. M1993, and by the Slovenian Research Agency (Grants No. P1-0403 and J1-2457).
	
	
		
\section{Data Availability}
Source data are available in the main text.

\bibliographystyle{model1-num-names}
\bibliographystyle{unsrt}
\bibliography{main}

\end{document}